\documentclass[sigconf]{acmart}

\usepackage{booktabs}
\usepackage{graphicx}
\usepackage{xcolor}
\usepackage{tikz}

\usetikzlibrary{positioning, shapes.geometric, arrows.meta, fit, backgrounds}

\copyrightyear{2026}
\acmYear{2026}
\setcopyright{cc}
\setcctype{by}
\acmConference[MODELS 2026]{ACM/IEEE 29th International Conference on Model Driven Engineering Languages and Systems}{October 04--09, 2026}{Málaga, Spain}
\acmBooktitle{ACM/IEEE 29th International Conference on Model Driven Engineering Languages and Systems (MODELS 2026), October 04--09, 2026, Málaga, Spain}
\acmDOI{10.1145/3822455.3838783}
\acmISBN{979-8-4007-2809-9/2026/10}

\begin{document}

\title{\texorpdfstring{Models as Governed Interfaces for AI-Native MBSE:\\
       Read-Side Adequacy and Write-Side Admissibility}{Models as Governed
       Interfaces for AI-Native MBSE: Read-Side Adequacy and Write-Side
       Admissibility}}

\author{Jason D. Gower}
\orcid{0009-0001-5945-6294}
\affiliation{%
  \institution{Loughborough University}
  \city{Loughborough}
  \country{United Kingdom}}
\email{J.Gower@lboro.ac.uk}

\author{Michael J. de C. Henshaw}
\orcid{0000-0003-0511-179X}
\affiliation{%
  \institution{Loughborough University}
  \city{Loughborough}
  \country{United Kingdom}}
\email{M.J.d.Henshaw@lboro.ac.uk}

\author{Siyuan Ji}
\orcid{0000-0001-6139-3539}
\affiliation{%
  \institution{Loughborough University}
  \city{Loughborough}
  \country{United Kingdom}}
\email{S.Ji@lboro.ac.uk}



% ---------------------------------------------------------------

\begin{abstract}
Programmatic access to machine-readable models such as SysML~v2 is often treated as sufficient for AI participation in systems engineering. It is only a precondition. The remaining work lies in the data architecture around the model. An AI reader querying a structurally complete model encounters absent derivation chains, untagged epistemic status, missing provenance, and unreachable evidence. Faced with these gaps, the AI reader fills them from training data, which lies outside the governed record. We test on the public Apollo~11 SysML~v2 reconstruction, exemplary by current practice. We name the missing property epistemic adequacy and offer it as a candidate data-architecture pattern in two halves. Read-side adequacy lets derivation, status, and provenance answer a query the model would otherwise leave to inference; write-side admissibility gates an AI contribution before it enters the record. The property decomposes into five criteria. Four sit on the read side, evidenced by the case and convergent literature; the fifth sits on the participation side, advanced as a hypothesis for future test. The architecture space spans an inline metadata extension to a substrate-native multi-model store, and over that space we propose the Governed-Query Architecture Framework, governing agent participation through the viewpoint conventions engineers already use. We state the reframing in falsifiable form: the epistemic layer stands only if it outperforms a retrieval-augmented baseline on the same model.
\end{abstract}

\maketitle

% ================================================================
\section{Introduction}\label{sec:intro}
% ================================================================

\looseness=-1
AI integration in model-based systems engineering has moved from concept
to practice: INCOSE's SE Vision~2035 names AI-augmented engineering a
strategic enabler~\cite{INCOSE2021Vision2035}, and deployed tooling spans
natural-language interfaces, retrieval-augmented generation, and
AI-assisted requirements
analysis~\cite{Zhang2026MBSECoPilot}. SysML~v2
extends this trend. Its machine-readable textual syntax, the
Kernel Modeling Language metamodel, and programmatic access through the
Systems Modeling API (SMAPI) make the model programmatically addressable
for the first time~\cite{OMGSysMLv2}. Machine readability is
necessary for AI engagement. An AI acts toward
the model in one of two roles~\cite{Ji2026ConsumerParticipant}. A
\textit{consumer} reads the model to answer a question. A \textit{participant}
proposes changes that the substrate must capture and
govern. A syntactically valid model
can still omit the derivation chains, epistemic-status declarations,
provenance, and reachable evidence an AI consumer needs to judge whether
a claim is grounded. Absent that metadata, the consumer fills
the gap from training data, which lies outside the governed record.
Retrieval reduces this substitution and leaves a residue of
it~\cite{Magesh2025HallucinationFree}.

\looseness=-1
We characterise these omissions as failures of
\textit{epistemic adequacy}. Epistemic here carries its standard sense: it
concerns how a claim is known rather than what the claim asserts. An
epistemically adequate substrate therefore records whether a value is
derived, at what standing, from what source, and against what
evidence. The failures divide
into four consumption-side gaps that convergent published evidence
confirms (derivation, status, provenance, and substrate completeness), and
one participation-side hypothesis on contribution governance that still
awaits test.
Closing the gap is an architecture question. Epistemic adequacy is a
candidate cross-domain data-architecture pattern for governed AI
participation. The response in MBSE is a co-design problem across
language metadata, methodology, and tool
governance~\cite{Ji2026ConsumerParticipant}.

\looseness=-1
We state the cross-domain claim as an open
hypothesis. The substrate is the model plus any adjacent
evidence layer the AI can reach as one queryable whole. Wherever evidence
is synthesised into an authoritative artefact with AI participation, we
conjecture that the same structural requirements hold.
Governed AI participation denotes an AI that consumes or contributes
under epistemic-adequacy constraints. The
word governed is deliberate: the architecture
guarantees only that every entry reaching the record is structurally
governed, and semantic correctness lies outside that guarantee.
Adjacent domains plausibly share the pattern: clinical decision
support and regulatory assembly arrive independently at provenance, status,
and derivation requirements~\cite{AhmedKirschner2026ProvenanceGap}. None of them, however, supplies review-gated
promotion into an authoritative record.

Throughout, the model is recast as the interface through which an AI
participant reasons about engineering constraints under
governance.
This paper makes three contributions. First, the paper reframes epistemic
adequacy as a data-architecture pattern, read-side adequacy paired with
write-side admissibility. A
language fixes what is expressible. What a given model populates, and what
its write path refuses, are properties of the deployment.
Second, the paper proposes what the write-side layer should consist of: eight admissibility
constraints, each built from an established governance mechanism and novel
only in conjunction, that together define the Governed-Query Architecture Framework
(GQAF), whose viewpoints, model kinds, and
correspondence rules govern agent participation the same way they
structure human comprehension. Third, the paper lays out a staged architecture
space, Candidates A, B, and C, recommends the metadata graph as the
deployable start and the substrate-native store as the
research programme's build target, and provides a falsifiable programme testing the
epistemic layer against retrieval-only access on the same model and prompt.

% ================================================================
\section{The Gap}\label{sec:gap}
% ================================================================

\subsection{The Apollo~11 Vignette}

\looseness=-1
We test the claim with a probe over the
real, public Apollo~11 SysML~v2
reconstruction~\cite{Helle2026Apollo11SysMLv2}, paired with a fictional
out-of-distribution control, so training-recalled answers are separate from
invented ones~\cite{ApolloConsumptionProbe2026}.
The model is structurally
complete: a requirement asserts five F-1 engines on the S-IC stage, with
free-text rationale ("necessary thrust to lift the massive Saturn~V
vehicle off the launch pad"), a stage-level minimum liftoff thrust
requirement of 34.5~MN, and a part definition carrying a per-engine
sea-level F-1 thrust of 6770~kN. Asked why five F-1 engines rather
than four or six, an LLM consumer working from the model alone finds no
governed answer. The machine-traversable, status-tagged derivation linking
the count to the thrust requirement is absent, and both values lack margin,
trade rationale, provenance, and epistemic status. The constituent values
are present; the authorising chain is absent. The
unlinked numbers conflict: five times 6770~kN is 33.85~MN, short of
the stated 34.5~MN minimum, and nothing in the model computes or flags the
discrepancy. Closing the chain needs
two external sources. The per-engine figure comes
from the Saturn~V AS-506 flight evaluation
report~\cite{NASA1969TMX58058}, and the four-versus-five engine
trade from the March~1961 Rosen committee paper
trail~\cite[p.~156]{Bilstein1980SP4206}. Both lie outside the model's reach.

Our experiments further put fifteen derivation-style questions to three current
commercial LLMs
over that model. Deadline-style pressure converted nine of
fifteen answers into content the substrate could not authorise. The
strongest model confabulated most fluently, and every invented claim was
historically plausible. A mocked EA1--EA4 metadata layer (Table~\ref{tab:criteria}) left a single ungrounded answer
across the forty-five trials under a governed
instruction and
halved the pressured failures~\cite{ApolloConsumptionProbe2026}. That layer is a hand-authored
sidecar over a pinned commit of the public model,
and the write-side constraints of \S\ref{sec:admissibility} remain
specified only; what the probe measures
is read-side consumption alone. Apollo is strongly
in-distribution, which is what makes it a demanding
test. A consumer can recover
the correct rationale from training and still fail the governance test,
because nothing in the substrate authorises the answer. A fabrication
that happens to be true reads as competence in review. The case
illustrates the property without establishing its frequency; its gaps are
what a well-formed model from competent practice tends to omit.

The criteria and constraints governing the probe were iteratively drafted and adversarially critiqued through interactions with Claude Opus 4.8. Each resulting criterion was independently validated against published literature (\S\ref{subsec:converge}). The authors manually verified every claim and assume full responsibility for the methodology and outputs.

\subsection{Convergent Literature}\label{subsec:converge}

\looseness=-1
Independent published work converges on the same diagnosis. Graph-based
link prediction recovers missing structural relationships in SysML models
at 72\% accuracy on a single aerospace case study,
direct evidence of missing structure in a
model of that kind~\cite{Karagoz2026MissingKnowledge}.
Decision-capture research at Airbus documents that rationale, evidence, and
justification are scattered across meetings, email, and tacit knowledge,
structurally absent from the model itself~\cite{Selmi2026DecisionMBSE}. SysML~v2 carries no native uncertainty
constructs, a gap that proposed extensions set out to
close~\cite{Zhang2026UncertaintySysML}. The UML Profile MARTE, by contrast, marks
whether a value is required, estimated, calculated, or measured~\cite{OMGMARTE}
without obliging any model to record it. Retrieval-augmented pipelines over
MBSE
are motivated by the insufficiency of base model content ~\cite{Esho2026RAGMBSE}, and adding structured dependency types lifts
LLM precision on mechatronic test-case generation from 0.41 to
0.76~\cite{May2026BoundaryConditions}. Surveys of LLM use in model-driven
engineering report applications clustered on natural-language interfaces and
content generation, with governance and evidence traceability largely
unaddressed~\cite{Rubei2025LLMsMDE}.
The diagnosis is consistent across these sources. The remainder of the paper
specifies our position on how the issue could be addressed architecturally. %the architectural response it licenses.

% ================================================================
\section{The Epistemic Adequacy Criteria}\label{sec:framework}
% ================================================================

A substrate is epistemically adequate when the groundedness of any claim
it holds is machine-decidable by query: whether the claim is derived, at
what standing, from what origin, with its premises reachable, and how it
entered the record. The property decomposes into five criteria, EA1--EA5
(Table~\ref{tab:criteria}).
Four are on the consumption side, where an AI reads and must judge
whether a claim is grounded; one is on the participation side,
where an AI proposes and the substrate must record how the claim entered.
The first four close by exposing metadata, the last by governing writes
(Figure~\ref{fig:layers}).
We expect that dropping any one criterion leaves a
class of ungrounded answers
indistinguishable from grounded ones. The set fixes standing at the
moment of query, a deliberate temporal boundary: a value satisfying every
criterion can become stale once an upstream element is revised; obligatory
suspect-marking under revision is the first extension the framework
needs.

\begin{table}[t]
\caption{The five epistemic adequacy criteria and the gap each leaves open. EA1--EA4 are consumption-side gaps exhibited by the Apollo~11 case and corroborated independently; EA5 is a participation-side hypothesis awaiting test.}
\label{tab:criteria}
\scriptsize
\begin{tabular}{@{}l p{2.1cm} p{2.75cm} p{1.3cm}@{}}
\toprule
\textbf{ID} & \textbf{Criterion} & \textbf{Gap left open} & \textbf{Support} \\
\midrule
EA1 & Derivation \& rationale & the requirement-to-value chain, and the alternatives weighed, is absent or trapped in free text & \cite{Selmi2026DecisionMBSE} \\
EA2 & Epistemic status & target, verified value, assumption, and placeholder are represented identically & \cite{Zhang2026UncertaintySysML} \\
EA3 & Provenance & who produced a value, by what method, at what stage, against what evidence is unrecorded & \cite{W3CPROV:2013} \\
EA4 & Substrate completeness \& reachability & values needed to close a chain lie outside the model or cannot be resolved & \cite{Esho2026RAGMBSE,Karagoz2026MissingKnowledge} \\
\midrule
\multicolumn{4}{@{}l@{}}{\textit{Participation-side hypothesis (not demonstrated by the case):}} \\
EA5 & Contribution governance \textbf{[H]} & nothing marks a claim's AI origin or routes it through review before it enters the record & \cite{NISTSP80053r5,EUAIAct2024} \\
\bottomrule
\end{tabular}
\end{table}

\textbf{EA1: derivation and rationale.} The logical chain from a
requirement to a value, including alternatives weighed and margins applied,
is either absent or recorded as free text an AI cannot traverse. The consumer
reads the value; the reasoning behind it stays out of reach, and an
unconstrained LLM fills the chain from training. Decision-capture
research finds this the norm, with rationale scattered outside the
model~\cite{Selmi2026DecisionMBSE}.

\textbf{EA2: epistemic status.} Design targets, verified values,
assumptions, and placeholders are represented identically, so a consumer
cannot separate a number confirmed by analysis from a working guess.
SysML~v2's verification cases return pass/fail verdicts on requirements
under test; no native construct marks the standing of an individual value,
a gap that proposed uncertainty extensions set out to
close~\cite{Zhang2026UncertaintySysML}.

\textbf{EA3: provenance.} Who produced a value, by what method, at what
review stage, and against what evidence goes unrecorded at the model
level. Provenance separates a value an auditor can trace to a responsible
source from one of unrecorded origin; W3C PROV supplies the vocabulary,
yet MBSE substrates rarely carry it on individual elements~\cite{W3CPROV:2013}.
Status says how settled a value is; provenance says where it came from;
a trustworthy answer needs both.

\textbf{EA4: substrate completeness and reachability.} The values needed
to close a derivation chain often live outside the
model, in evidence
beyond the AI's reach or behind a source reference that fails to
dereference. In the Apollo exemplar, both closing values are real and
unreachable. Retrieval pipelines over MBSE respond to this
shortfall~\cite{Esho2026RAGMBSE,Karagoz2026MissingKnowledge}.

\looseness=-1
\textbf{EA5: contribution governance.} When an AI proposes a claim,
nothing in the substrate marks its origin or routes it through review
before it enters the authoritative record. The gap differs from the other
four in evidentiary status: EA1--EA4 are demonstrated by the Apollo case
and the literature in \S\ref{subsec:converge}; EA5 is a
hypothesis this paper advances
and leaves for future test. EA5 belongs
with the others for a structural reason. An architecture that closed
EA1--EA4 and left writes ungoverned would surface grounded answers while
ungrounded contributions accumulated unchecked. The remedy is structural:
the substrate decides what enters the record
and under what review.

% ================================================================
\section{The Admissibility Constraints}\label{sec:admissibility}
% ================================================================

\begin{figure}[t]
\centering
\includegraphics[width=1\columnwidth]{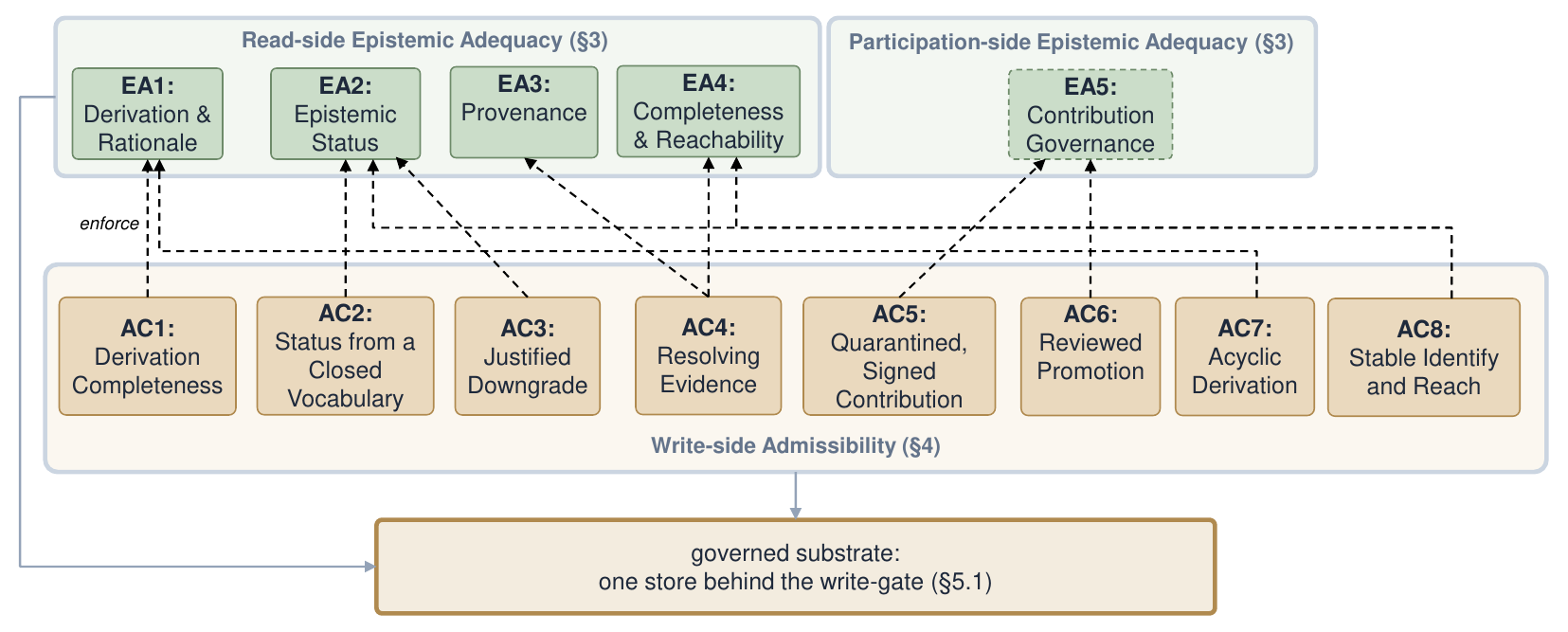}
\caption{Three layers: criteria fix what the substrate exposes,
constraints what it refuses, one store implements both; dashed
participation-side EA5 is enforced by AC5 and~AC6.}
\label{fig:layers}
\Description{Three stacked layers: a green read-side adequacy box
containing chips EA1 to EA4, with a dashed participation-side EA5 chip
beside it; an orange write-side admissibility box containing chips AC1 to
AC8; and an orange governed substrate slab beneath. Arrows labelled
enforce run upward from the AC5 and AC6 chips, converging on the dashed
EA5 chip.}
\end{figure}

\looseness=-1
The criteria in \S\ref{sec:framework} govern what an adequate substrate exposes; this section governs what it refuses. An architecture could surface
derivation, status, and provenance for the values already in the record
and still let an ungrounded value enter unchallenged, so the criteria need
a companion layer that governs writes; Figure~\ref{fig:layers} places the
two layers and the store that implements them. We state the admissibility
layer as eight constraints, AC1--AC8, each blocking a write unless a
required property holds. Read together, the eight are integrity
constraints over the model substrate~\cite{Codd1970RelationalModel}: each
refuses a write that would violate a stated property. A transaction's
consistency property admits only constraint-satisfying
states~\cite{HaerderReuter1983ACID}, and an Object Constraint Language
invariant rejects a breaching model instance~\cite{OMG:OCL24:2014}.
The constraints are architecture-independent, fixing what any candidate
solution (see \S\ref{sec:candidates}) must enforce and leaving the mechanism open; each enforces one or more of the criteria, as Table~\ref{tab:admissibility} records. The
parenthetical in each heading below and in that table names the criterion
the constraint enforces.

\textbf{AC1: derivation completeness (EA1).} A governed value must reference
at least one upstream element or evidence source, and a governed number
must carry either a derivation expression or an explicit unresolved marker.
The marker separates a value awaiting derivation from one whose
basis was never recorded: an open chain must declare itself
open. Without the constraint a value sits with no traceable
basis, the orphaned-value condition that requirements-traceability research
finds dominant~\cite{Gotel1994Traceability}, and an unconstrained AI
fills the missing chain from training.

\textbf{AC2: status from a closed vocabulary (EA2).} The epistemic status of
a value is drawn from a fixed set of terms, and the field is mandatory on
every governed value, so that a design target, a working assumption, and a verified
result stay distinguishable to a machine reading the model.
Drawing status from an enumerated domain
is a domain constraint in the relational
sense~\cite{Codd1970RelationalModel}, checkable as a model invariant at
the moment of writing~\cite{OMG:OCL24:2014}. MARTE makes the same
distinction expressible~\cite{OMGMARTE}; AC2 makes it obligatory.

\textbf{AC3: justified downgrade (EA2).} A change that lowers a value's
standing, say from verified to assumption, is admitted only with a
justification record naming author, date, and reason. Downgrades are
governed more strictly than other edits because they are the failure most
likely to pass unnoticed: weakening a value erodes the evidentiary basis
of everything derived from it, and the record turns that weakening into an
accountable event.

\textbf{AC4: resolving evidence for safety-critical elements (EA3, EA4).} A
safety-critical element commits only with complete provenance, the agent,
method, and review stage behind its value, together with an evidence URI
that resolves.
The constraint demands a link that
dereferences at commit, an admissibility check rather than a durability
guarantee. The full provenance burden falls only on safety-critical
elements, where the cost of an unsupported value is
highest~\cite{Gotel1994Traceability,W3CPROV:2013}.

\begin{figure}[t]
\centering
\includegraphics[width=\columnwidth]{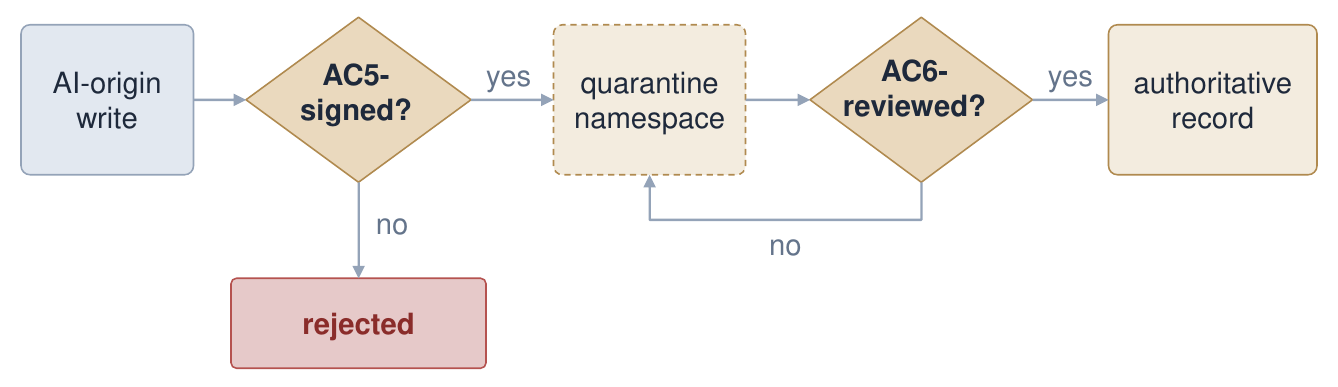}
\caption{The AC5/AC6 write lifecycle: unsigned writes are rejected;
signed ones wait in quarantine until a second-party review record
(reviewer $\neq$ proposer) promotes them; review depth tiers by
criticality.}
\label{fig:lifecycle}
\Description{A left-to-right state diagram: an AI-origin write reaches an
AC5 signed decision diamond, where unsigned writes are rejected; signed
writes enter a dashed quarantine namespace, then an AC6 review decision
diamond, whose no branch loops back into quarantine and whose yes branch
promotes the write into the authoritative record.}
\end{figure}

\textbf{AC5: quarantined, signed contribution (EA5).} An unsigned AI-origin
write is rejected outright, and a signed one enters a quarantine
partition (Figure~\ref{fig:lifecycle}). Attribution is the precondition for governing contribution:
a record that cannot say which agent produced a value cannot route it
for review, and holding AI proposals apart until reviewed is the
separation-of-duties posture high-assurance controls require for any
privileged write~\cite{NISTSP80053r5}.

\looseness=-1
\textbf{AC6: reviewed promotion (EA5).} Promotion out of quarantine
requires a completed review record giving the reviewer, the disposition,
and the date. The reviewer must differ from the proposing agent, so no principal
both authors and admits its own contribution; the record is the auditable
evidence that a responsible human approved it. This is the dual-authorisation
pattern applied to AI participation: admission turns on a recorded
second-party
review~\cite{NISTSP80053r5}. Review depth follows the criticality scoping
AC4 already uses: a safety-critical promotion requires the full
second-party record; lower-criticality promotions may enter under a
standing policy record with sampled human audit. This concentrates review
effort where an unsupported value costs most. AC5 and AC6 realise the
participation-side EA5 and inherit its status: both remain proposals
awaiting experimental validation.

\textbf{AC7: acyclic derivation (EA1).} A write that would make the
derivation graph cyclic is refused at commit. A value that
depends on itself, directly or through intermediaries, has no base
case; build systems have long
taken the same posture toward circular dependencies.

\textbf{AC8: stable identity and reach (EA2, EA4).} Each sideways
identifier (a cross-reference into the \S\ref{sec:candidates}
metadata graph) must resolve to exactly one live element and one graph node per revision,
and conflicting claims or statuses on the same element are blocked. An
identifier that resolves to nothing, or to several elements at once,
breaks the join an AI consumer relies on to connect a value to its
derivation and status. This is primary-key uniqueness and foreign-key
reference~\cite{Codd1970RelationalModel}, recast for a dereferenceable
substrate.

\looseness=-1
Each constraint
is enforced by blocking the write at commit, the posture
the EU AI Act reflects in requiring high-risk systems
to log events automatically and to govern the data they depend
on~\cite{EUAIAct2024}. The layer's reach
is bounded by design: it guarantees
only the absence of
structurally ungoverned entries.
Rubber-stamped reviews, coordinated falsification, signing-key misuse, and
broken-but-resolvable URIs all sit outside that reach. A human reviewer
therefore remains the terminal author of any safety-relevant
claim~\cite{NSPE2023ResponsibleCharge}.
\looseness=-1
The authoring cost deserves acknowledgement. Traceability regimes have
historically failed on cost, because the burden of supplying trace data
falls on people other than its beneficiaries~\cite{Gotel1994Traceability}.
A blocking gate raises the
cost of every write. Two features
limit the burden. First, AC1's floor is low: a write satisfies AC1 by declaring an open evidence chain, nothing more. Second,
on AI-origin writes the proposing agent drafts its
own governance metadata (\S\ref{subsec:srset}), placing the cost on
software for exactly the write class the gate exists to govern. One
failure mode still survives, a substrate in which unresolved markers become the norm; \S\ref{sec:future} therefore tracks the
unresolved-marker rate.

\looseness=-1
\textbf{Relation to existing mechanisms.} Every mechanism below already exists, and none is
our contribution. What none of them binds together is the conjunction this
paper claims: constraints that are epistemic rather than structural, held
per value, applied to writes typed by AI origin,
and enforced at the gate rather than audited afterwards. Each supplies one
element of the conjunction and lacks the rest. Provenance vocabularies give agent, method,
and activity~\cite{W3CPROV:2013}, and quarantined staging with dual
authorisation is a standard high-assurance control~\cite{NISTSP80053r5}.
The UML Profile for MARTE provides a reference to read-side vocabulary, marking a value as required, estimated, calculated, or measured~\cite{OMGMARTE}; the mark is
available to a model rather than required of one, and no write path
enforces it. Configuration management contributes closed maturity vocabularies and
second-party release records, but governs the release rather than the
value~\cite{ISO10007}. Assurance cases give a machine-readable claim,
evidence, and status structure, but only once the claim has been
made~\cite{OMGSACM}. Splitting the property into read-side criteria and
write-side constraints
makes the conjunction checkable. We offer that
split, and the architectures that carry it,
for falsification.

\begin{table}[t]
\caption{The eight admissibility constraints, the property each enforces, and its supporting literature.}
\label{tab:admissibility}
\scriptsize
\begin{tabular}{@{}l p{4.7cm} p{1.15cm}@{}}
\toprule
\textbf{Constraint} & \textbf{The write is blocked unless\ldots} & \textbf{Support} \\
\midrule
AC1 (EA1) & a derivation references $\geq 1$ upstream element or evidence reference, and a governed numeric value carries a derivation expression or an explicit unresolved marker & \cite{Gotel1994Traceability,Codd1970RelationalModel} \\
AC2 (EA2) & the epistemic status is drawn from the closed vocabulary (no free text, no omission) & \cite{Codd1970RelationalModel,OMG:OCL24:2014, OMGMARTE} \\
AC3 (EA2) & a status downgrade carries a justification record (author, date, reason) & \cite{W3CPROV:2013,HaerderReuter1983ACID} \\
AC4 (EA3/EA4) & a safety-critical element has complete provenance (agent, method, stage, and a resolving evidence URI) & \cite{Gotel1994Traceability,W3CPROV:2013} \\
AC5 (EA5) & an AI-origin value enters only quarantined; unsigned writes are rejected & \cite{NISTSP80053r5,EUAIAct2024} \\
AC6 (EA5) & promotion from quarantine carries a completed review record (reviewer, disposition, date); review depth tiers by criticality & \cite{NISTSP80053r5} \\
AC7 (EA1) & the resulting derivation graph stays acyclic & \cite{OMG:OCL24:2014} \\
AC8 (EA2/EA4) & each sideways identifier resolves to exactly one live element and one graph node per revision; conflicting claims or statuses are blocked & \cite{Codd1970RelationalModel} \\
\bottomrule
\end{tabular}
\end{table}

% ================================================================
\section{Candidate Architectures}\label{sec:candidates}
% ================================================================

\begin{figure*}[t]
\centering
\includegraphics[width=0.95\textwidth]{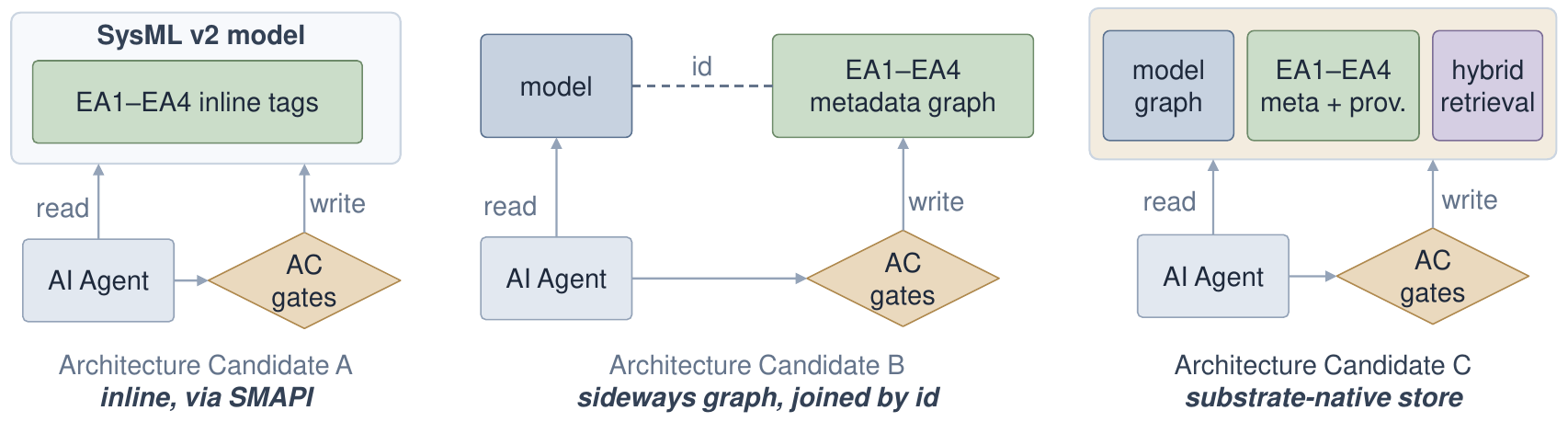}
\caption{Candidate architectures in order of increasing internalisation;
colour tracks capability (model blue, metadata green, index violet, store
orange). The AC1--AC8 gate sits on every write path; only its placement
moves.}
\label{fig:candidates}
\Description{Three side-by-side schematics of candidate architectures A, B
and C, using one colour per capability: in A a green epistemic-metadata
block sits inside the blue model box; in B the green block stands beside
the blue model as a separate graph joined by identifier; in C blue model,
green metadata and a new violet retrieval index sit together inside one
orange store boundary. Each panel has an admissibility gate on the AI
write path.}
\end{figure*}

The three candidates (Figure~\ref{fig:candidates}) differ chiefly in how
much they ask the toolchain and the language to change, and they are
staged: a programme can begin at~A or~B and move
toward~C as AI participation matures. None of the three is built. Each is
described in the present tense of a design specification, and
\S\ref{sec:future} states the order in which we intend to test them.

\textbf{Candidate~A: inline metadata through an SMAPI extension.} EA1--EA4
metadata is carried directly on SysML~v2 elements through the standard
\texttt{metadata def} mechanism, accessed and mutated through SMAPI; EA4
reachability resolves as a query over the element graph plus URI-typed
evidence links. EA5 is
enforced at the SMAPI access layer as a write gate. The strength is archival durability: governance metadata
travels with the model artefact. The weaknesses are standards latency,
since OMG adoption and uniform vendor implementation gate cross-tool
interoperability, and the assumption that SMAPI is the sole write path;
vendor batch loaders or direct database writes bypass enforcement.

\textbf{Candidate~B: sideways-identifier metadata graph.} The
engineering semantics of the model stay unchanged. A separate metadata
graph, keyed by programme-stable element identifiers, carries EA1--EA4.
An AI consumer reads both artefacts and joins them on the
identifier. An AI participant submits a proposed model change together
with a graph entry, and a governance gate evaluates EA5 on the graph
side before the model write proceeds. EA4 becomes a graph query surfacing broken
links and missing nodes as first-class findings. The
shape has precedents. Named graphs supply the sideways-metadata
axis~\cite{Carroll2005NamedGraphs}, OSLC federates metadata across
tools~\cite{OSLCCore2021}, and view-based consistency preservation repairs
structural drift~\cite{Klare2021Vitruvius}. All three leave AI-origin
writes ungoverned. What~B adds to those three is
the gate of \S\ref{sec:admissibility}. The cost is synchronisation risk: model and
graph are two systems of record, and drift between them is the standing
operational failure mode.

\textit{Candidate~B assumptions.} B requires programme-stable identifiers
and mediated writes. Each graph node carries a content hash of the
element state it describes, so a native-tool value edit under a stable
identifier surfaces as divergence at the next read. Identifier
stability across tool round-trips is a research risk the pilot
reports on.

\textbf{Candidate~C: substrate-native multi-model store.} The most
ambitious option abandons the two-record split: one store
internalises the model topology, the EA1--EA4 metadata and provenance, and
the AI-retrieval index as a single queryable whole, with AI agency
governed inside the store rather than at its perimeter. Because model graph and
governance metadata commit under one transactional engine, B's two-record
drift is reduced within the governed write path; the retrieval index
becomes a derived, eventually-consistent replica
rather than a third
system of record. C is less a
deployable product than a target: the store it asks for is the vision
of~\S\ref{subsec:srset}. Its costs are higher: it needs an SMAPI-class
connector as the sole write path and concentrates trust in one engine, a
larger failure surface than~B's split design.

B is the strongest deployable starting point. It works with today's
standards, enforces the minimum read-side metadata for
EA1--EA4, gates EA5, and exposes EA4 gaps as first-class graph queries.
B's gate also owns the one artefact it
guards, the graph itself. A's
metadata, by contrast, lives inside the vendor-managed model, where any
native write can reach it. B is also the natural
staging ground for~C: the identifier-keyed graph B maintains is the
schema C internalises as a native store,
with the model topology added. Candidate~C is the programme's build target, evaluated
against A and~B in \S\ref{sec:future}.

% ================================================================
\section{Vision: The Governed Substrate and Agentic Workflows}\label{sec:vision}
% ================================================================

Once MBSE models become LLM-manageable, the model is the layer
through which AI engages physical-world constraints under governance;
\S\ref{subsec:srset} states the requirements Candidate~C's substrate must
satisfy to realise that role today.

\subsection{Realising the Substrate}\label{subsec:srset}

Epistemic adequacy is achievable ahead of any language standard.
The vision asks for one store holding model graph, governance metadata,
and retrieval index behind a transactional write-gate that evaluates the
admissibility constraints, enforces status as a closed vocabulary, and
keeps provenance append-only, a history that also makes expressible the
suspect-marking under revision that \S\ref{sec:framework} defers.
A hypothetical next-generation language, referred to here as
SysML~v3, would
standardise the matching primitives, epistemic-status keywords and a
derived-from construct among them; met by such a store, those primitives become
enforceable at the substrate.

Engines combining property-graph, document, and vector storage are an
established category~\cite{Coimbra2025GraphDatabases}; the storage
primitives exist today. A deployment must still add serialisable commit
over the authoritative graph and document state, keeping the retrieval
index a derived replica behind the write-gate. Such a store would
internalise AI-agency governance, holding the authoritative record and an
AI-authored companion as two namespaces of one engine. An AI
participant writes EA1--EA5-tagged proposals to the companion namespace, and
the write-gate promotes accepted content in a single transaction
(Figure~\ref{fig:lifecycle}). A
SMAPI-class connector as the sole write path remains a precondition the
organisation enforces. The exemplar engines still lack the qualification
record and serialisable isolation that safety-critical adoption demands.

\subsection{Agentic Workflows Within Workflows}

A conventional MBSE workflow is a lifecycle-structured sequence of
human-directed activities. Agentic participation inserts AI sub-workflows
bounded by viewpoint and handoff scope, as engineer-supervised AI over
toolchains already does~\cite{Son2026AgenticWorkflow}. A
retrieval agent answers a query, a synthesis agent assembles a
safety-relevant view, and a governance agent tags AI contributions with
origin and status before a reviewer sees them. The
probe in \S\ref{sec:gap} found the retrieval and synthesis roles
confabulating under pressure, so the safety argument rests on the
governance role and the substrate's
refusal.

\subsection{The Governed-Query Architecture Framework}\label{subsec:gqaf}

\begin{figure}[t]
\centering
\includegraphics[width=\columnwidth]{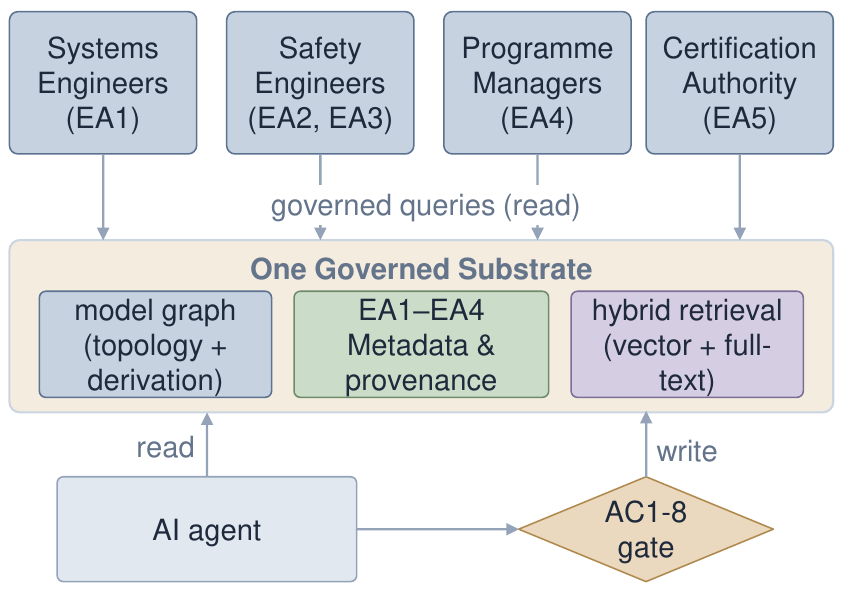}
\caption{The Governed-Query Architecture Framework (GQAF): every
stakeholder viewpoint, and every AI agent extraction, resolves to a
governed query over one substrate; AI contributions enter only through
the AC1--AC8 gate.}
\label{fig:gqaf}
\Description{Four stakeholder viewpoints (systems engineer, safety
engineer, programme manager, certification authority) issue governed read
queries to one shared governed substrate, drawn as an orange boundary
containing a blue model graph, green metadata and provenance, and a violet
retrieval index; an AI agent likewise reads by governed query and
contributes only upward through an AC1--AC8 admissibility gate.}
\end{figure}

Stakeholders and agents query the same substrate along different traversal
paths (Figure~\ref{fig:gqaf}); epistemic adequacy therefore has to hold at
the substrate, where all those paths meet.

We
call the synthesis an Architecture Framework (AF) for Agent-Driven
Development. The viewpoints, model kinds, and correspondence rules an AF
already provides to structure human comprehension become the same
mechanisms that scope what AI participants can read and
write. Every viewpoint resolves to
a governed query over one substrate because the AF's read-side conventions
already imply that convention~\cite{Atkinson2010OSM}; in
ISO~42010 terms~\cite{ISOIEC42010} a viewpoint is a query specification
and a view is its governed result. What the AF adds is write-side
admissibility over those same queries, the AC1--AC8 gate in
\S\ref{sec:admissibility}. We name the resulting pattern the
\textit{Governed-Query Architecture Framework} (GQAF).

The substrate in \S\ref{subsec:srset}, the agentic workflows above, and
the candidates in \S\ref{sec:candidates} are one architecture seen from
three sides.
One known obstacle remains unresolved by the
framework. An agent whose epistemic scope is one
viewpoint commits base-level
writes. Those writes can conflict with other
viewpoints' concerns in ways the structural checks cannot see: the
view-update problem in its
governed form~\cite{BancilhonSpyratos1981}. GQAF bounds rather than solves
it. Every AI-origin write lands in quarantine (AC5) and is admitted only
on second-party review (AC6). Cross-viewpoint conflicts that structure
cannot detect are routed to the human reviewer, which is what makes that
review a structural component of the framework.

% ================================================================
\section{Evaluation Plan and Conclusion}\label{sec:future}
% ================================================================

The next step is to investigate the architectures through action design
research~\cite{Sein2011}. The Apollo~11 thrust chain (\S\ref{sec:gap}) is the
first probe; its read-side run already halves pressure-induced ungrounded
answers, with the gain appearing as correct abstention~\cite{ApolloConsumptionProbe2026}.
Later iterations extend to further MBSE subsystems.

Each iteration asks whether epistemic metadata, beyond structured access
alone, lets an AI answer multi-hop derivation queries more reliably. Three
arms under a fixed model and prompt: retrieval over unstructured content, the
structured model stripped of epistemic metadata, and the structured model
carrying it. Structure already lifts LLM precision on model
content~\cite{May2026BoundaryConditions}, so that second arm is our null model; the claim
at risk falls between the second and third arms. Groundedness~$G$ and Correct
Abstention~$A$ adapt RAG faithfulness~\cite{Es2024RAGAS} and
selective-answering scoring~\cite{Kamath2020SelectiveQA}, and
Derivation-chain Completeness~$D$ extends multi-hop supporting-fact recovery
to ordered chains. Contribution Governance Fidelity~$F$ is new, counting
promotions without a review record and mis-attributed origins. Promotions whose review
record is present but perfunctory lie outside the metric. $A$ and~$F$ discriminate: prior structure results say
nothing about when to refuse or how faithfully a contribution's history is
recorded. Two gate-produced failure modes are also scored: the
unresolved-marker rate and a spot-audit of admitted AI-origin derivations. If
the substrate cannot outperform the stripped arm on $A$ and~$F$ across two cycles,
we judge the claim refuted.

Implementation proceeds in stages. We build the Apollo chain inside a
Candidate~C-style store, Candidate~B as fallback, then test AC1--AC8 as write
gates and GQAF viewpoints as governed queries. A live-project trial follows. The trial
provides supporting evidence if a review deliverable resolves its evidence
from the substrate and the gate blocks a contribution that violates a
constraint.

\looseness=-1
Our position is that what closes the gap is data architecture rather than a richer language: the
model becomes the governed interface through which AI participates, and the
claim stands only if the epistemic layer outperforms the stripped arm.

\bibliographystyle{ACM-Reference-Format}

\bibliography{refs}

\end{document}